\documentclass[prb,aps,twocolumn]{revtex4}
\usepackage{amsmath,bm}
\usepackage{graphicx}

\newcommand{\be}{\begin{equation}}
\newcommand{\ee}{\end{equation}}
\newcommand{\BE}{\begin{eqnarray}}
\newcommand{\EE}{\end{eqnarray}}
\newcommand{\BEn}{\begin{eqnarray*}}
\newcommand{\EEn}{\end{eqnarray*}}
\newcommand{\barr}{\begin{array}} 
\newcommand{\earr}{\end{array}}

\newcommand{\bit}{\begin{itemize}}
\newcommand{\eit}{\end{itemize}}
\newcommand{\bfl}{\begin{flusleft}}
\newcommand{\efl}{\end{flusleft}}
\newcommand{\bfr}{\begin{flushright}}
\newcommand{\efr}{\end{flushright}}
\newcommand{\bfig}{\begin{figure}}
\newcommand{\efig}{\end{figure}}

\newcommand{\bc}{\begin{center}}
\newcommand{\ec}{\end{center}}

\newcommand{\ben}{\begin{enumerate}}    
\newcommand{\een}{\end{enumerate}}

\newcommand{\eps}{\varepsilon}
\newcommand{\de}{\partial}
\newcommand{\e}{\mbox{e}}

\begin{document}

\pagestyle{plain}

\title{Mean Field Approach for a Statistical Mechanical Model of Proteins}  

\author{Pierpaolo Bruscolini}
\affiliation{Dipartimento di Fisica \& INFM, Politecnico di Torino,
C.so Duca degli Abruzzi 24, I-10129 Torino, Italy}
\author{Fabio Cecconi}
\affiliation{Dipartimento di Fisica Universit\`a "La Sapienza" \& INFM
Unit\`a di Roma1, P.le A.~Moro 2, I-00185 Roma, Italy}

\begin{abstract}
We study the thermodynamical properties of a topology-based model 
proposed by Galzitskaya and Finkelstein for the description of protein 
folding. We devise and test three different mean-field approaches for
the model, that simplify the treatment without spoiling the description.
The validity of the model and its mean-field approximations 
is checked by applying them to the 
$\beta$-hairpin fragment of the immunoglobulin-binding protein (GB1) and
making a comparison with available experimental data and simulation results. 
Our results indicate that this model is a 
rather simple and reasonably good tool for interpreting  folding 
experimental data, provided the parameters of the model are carefully 
chosen. The mean-field approaches substantially recover all the
relevant exact results and represent reliable alternatives to the 
Monte Carlo simulations.  
\end{abstract}
\maketitle
\date{\today}

\section{Introduction}

The free-energy landscape of protein molecules represents the
key-information for understanding processes of biomolecular
self-organization such as folding. 
\cite{Fun1,Fun2,Fun3,Fun4}
The free-energy landscape, indeed, determines all observable properties of 
the folding process, ranging from protein stability to folding rates. 
\cite{ChanDill,MuSerr,Jackson,Fersht,KarpRev}
Unfortunately, for real proteins, sophisticated all-atom computational
methods fail to characterize the free-energy surface, since they are
currently limited to explore only few stages of the folding process.
As an alternative, one can argue that,  
taking into account all the complex details of chemical interactions is 
not necessary to understand how proteins fold into their native state.
Rather, elementary models incorporating the fundamental physics of folding,
while still leaving the calculation and simulations simple, can 
reproduce the general features of the free-energy landscape and
explain a number of experimental results.
This attitude, typical of a statistical mechanics approach, agrees
with the widely accepted view that ``a surprising simplicity
underlies folding'' (Baker\cite{Baker}).
In fact several experimental\cite{Riddle,Fersht2,Serrano,Marqusee} and 
theoretical 
studies\cite{Plaxco2,Miche1,Clementi,Clementi2} indicate the topology of 
protein native state as a determinant factor of folding. 
As examples, one can mention the fact that even 
heavy changes in the sequence that preserve the native state, have a
little effect on the folding rates.\cite{Chiti,Riddle2} 
Moreover, the latter are found to correlate to the average
contact order,\cite{Plaxco} which is a topological property of native state.
Finally, proteins with similar native state but low sequence
similarity often have similar transition state ensembles.\cite{Chiti,Plaxco2} 

Within this context, elementary 
models\cite{GoMacro,GoBiopo,MEnature,Gauss,AlmBak}
which correctly embody the native state topology and interactions, are 
believed to be useful in describing the energy landscape of real proteins.
In this paper, we study one of such topology-based models 
proposed by Galzitskaya and Finkelstein (GF),\cite{Finkel} 
which was developed to identify the folding nucleus and the 
transition state configurations of proteins. 
The model employs a  
free-energy function  with a reasonable
formulation of the conformational entropy, which is certainly
the most difficult contribution to describe.
The energetic term, instead, takes into account only native state 
attractive interactions.
In the original paper,\cite{Finkel} the model was combined with a dynamic
programming algorithm to search for transition states of various
proteins.  To reduce the computational cost of the search, two
kinds of approximations were introduced: the protein was regarded as
made up of ``chain links'' of 2-4 residues, that fold/unfold together;
besides, only configurations with up to three stretches of contiguous
native residues were considered in the search (``triple-sequence
approximation''). 
As shown in Ref.~\onlinecite{Brusco}, the effect of such assumptions is a 
drastic entropy reduction of the unfolded state and possibly of the 
transition state. This produces free energy profiles very different 
from the true ones, thus spoiling the evaluation of $\phi$-values. 

Here, we apply the model in a more general statistical mechanical 
philosophy: namely, we develop three different mean-field approaches of
increasing complexity, and compare their prediction with the exact
results, obtained by exhaustive enumeration of all the
configurations, in the case of a 16-residues-long peptide (C-terminal 
41-56 fragment of the streptococcal protein G-B1)\cite{seqhair}  which is 
known to fold, in isolation, to a $\beta$-hairpin 
structure.\cite{MEnature}
Our main goal here is to test the model against experimental findings 
and to test the mean-field predictions against the exact results. 
In the future we will use this knowledge to apply the appropriate 
mean-field approach to the case of real proteins, for which exhaustive
enumeration is unfeasible. 

The paper is organized as follows. In the next section, 
we present and describe the main features of the GF model.
In section II, we introduce and discuss three mean-field
approximations: the usual scheme, and two
other approaches stemming from the knowledge of the exact solution for
the Mu\~noz-Eaton model.\cite{Brusco}
In section III, we apply the model and its mean-field
approximations to study the folding transition of the 
$\beta$-hairpin and discuss our results. 
         
\section{Description of Galzitskaya-Finkelstein Model}
The GF model assumes a simple description of the polypeptide chain, where
residues can stay only in an ordered (native) or disordered
(non-native) state. Then, each micro-state of a protein with $L$
residues is encoded in a sequence of $L$ binary variables
${\bf s}= \{s_1,s_2,...,s_L\}$, ($s_i=\{0,1\}$).
When $s_i=1$ ($s_i=0$) the $i$-th residue is in its native (non-native)
conformation.
When all variables take the value $1$ the protein is correctly folded, 
whereas the random coil corresponds to all $0$'s.
Since each residue can be in one of the two states, ordered or disordered, 
the free energy landscape consists of $2^L$ configurations only. 
This drastic reduction of the number of available configurations  
represents, of course, a restrictive feature of the model, 
however, follows the same line of the well known 
Zimm-Bragg model\cite{ZB} widely employed to describe the helix to 
coil transition in heteropolymers. 

The effective Hamiltonian (indeed, a free-energy function) is 
\begin{equation}
H({\bf s}) = \eps \sum_{i<j}  \Delta_{ij}  s_i s_j - T S({\bf s})\,,
\label{eq:finkel}
\end{equation}
where $S({\bf s})$ is given by:
\begin{equation}
S({\bf s}) = R \left[ q \sum_{i=1}^{L} (1 - s_i) + S_{loop}({\bf s})\right]\,.
\label{eq:S}
\end{equation}
$R$ is the gas constant 
and $T$ the absolute temperature. 
The first term  in Eq.~(\ref{eq:finkel}) is the energy associated to native 
contact formation. Non native interactions are neglected:  this assumption,
that can be just tested {\it a posteriori},
is expected to hold  if, during the folding process, the progress
along the reaction coordinate is well depicted on the basis of 
the native contacts (that is, the reaction coordinate(s) must be
related to just the native contacts). Moreover, such progress must be slow  
with respect to all other motions, so that all non-native interaction
can be ``averaged-out'' when considering the folding process.\cite{KarpRev}
$\Delta_{ij}$ denotes the element $i$,$j$ of the contact matrix,
whose entries are the number of heavy-atom contacts between 
residues $i$ and $j$ in the native state.
Here we consider two amino-acids in contact, when there are at least two heavy 
atoms (one from aminoacids $i$ and one from $j$) separated by a distance 
less than $5$\AA. 
The matrix $\Delta$ embodies the geometrical properties of the
protein.
Notice that, in the spirit of considering the geometry more
relevant than the sequence details, every (heavy) atom-atom contact is
treated on equal footing: the chemical nature of the atoms is ignored, 
together with a  correct account for the different kind of interactions.

The second term in (\ref{eq:finkel}) is the conformational entropy
associated to the presence of unfolded regions along the chain,
and vanishes in the native state.
 
More precisely the first term in Eq.~(\ref{eq:S}) is 
a sort of ``internal'' entropy of the residues, that can be
attributed to the ordering of the main and side-chains' degrees of
freedom upon moving 
from the coil to the native state. Indeed, $q R$ represents
the entropic difference between the coil and the native
state of a single residue, as can be noticed by considering that in
the fully unfolded state  the first and last term vanish, and the
entropy is given by $q L R$. 

The quantity $R S_{loop}$ in Eq.~(\ref{eq:S}), instead,  
is the entropy pertaining to the disordered closed loops protruding 
from the globular native state;\cite{Finkel2} it reads:
\begin{equation}
S_{loop}({\bf s}) = \sum_{i<j} J(r_{ij}) s_i s_j \prod_{k=i+1}^{j-1}(1-s_k)\,.
\label{Eq:Sloop}
\end{equation}
According to Ref.~\onlinecite{Finkel}, we take:\cite{privcomm}
\begin{equation}
J(r_{ij}) = -\frac{5}{2} \ln|i-j| -
\frac{3}{4}\frac{r_{ij}^2-a^2}{A a |i-j|}\,.
\label{Eq:J}
\end{equation}
In this context a disordered loop is described by a strand of 
all ``0''s between two ``1''s: for instance the configuration 
$11000000111100011$ contains 
two loops involving $6$ and $3$ residues respectively. The product
in expression~(\ref{Eq:Sloop}) warrants that only uninterrupted 
sequences of ``0'' can contribute to the loop entropy.  
The configuration of a disordered loop going from residues
$(i+1)$ to $(j-1)$, with $i$ and $j$ in their native positions,
is assimilated to a gaussian chain of beads ($C_{\alpha}$ atoms)
with end-to-end distance $r_{ij}$, the latter being the distance between
C$_{\alpha}$ atoms of residues $i$ and $j$ in the native state. 
The parameters $a=3.8$ \AA~  and $A=20$ \AA~ are the average distance 
of consecutive $C_{\alpha}$'s along the chain and persistence length 
respectively. Other forms for $S_{loop}$ could  also be used
(see, e.g.\ Ref.~\onlinecite{AlmBak}); yet, here we are interested in
evaluating the original GF model and devising good mean-field 
approximations to it, and we will not discuss this subject any
further. The interested reader may refer to the
original articles \cite{Finkel, Finkel2} for a derivation of
Eq.~(\ref{Eq:J}). 

\section{Mean Field Approaches to the GF Model}
Mean field approach (MFA) is certainly the first attempt to investigate
the thermodynamical properties of complex systems, because it provides a 
qualitative picture of the phase diagram that in many  
cases is only partially modified by more accurate refinement of the
theory.
In its variational formulation, 
MFA, for a system with Hamiltonian $H$ and corresponding 
free-energy $F$, starts from the Bogoliubov-Feynman inequality
\begin{equation}
F \leq F_0 + \langle H - H_0 \rangle_0\,,
\label{eq:bogofey}
\end{equation}
where $H_0$ is a solvable trial Hamiltonian $F_0$ is the corresponding
free-energy, both depending on free parameters
${\bf x} = \{x_1\cdots x_L\}$ (variational parameters). Such parameters
have to be chosen to minimize the second member of
(\ref{eq:bogofey}) to get the minimal upper bound
of $F$ and accordingly its better approximation.
This method defines a variational free-energy
\begin{equation}
F_{var} = F_0 + \langle H - H_0 \rangle_0\,,
\label{Eq:genericFvar}
\end{equation}
whose minimization leads to the self consistent equations
that in their general form read
\begin{equation}
\bigg\langle \frac{\partial H_0}{\partial x_l} \bigg\rangle_0 
\langle H-H_0\rangle_0 - \bigg \langle (H-H_0)
\frac{\partial H_0}{\partial x_l}
\bigg\rangle_0 = 0\,, 
\label{Eq:meanfield}
\end{equation}
 with $l=1,\ldots, L$. We implement different versions of the MFA for
the GF model that differ each from the other 
by the choice of the trial Hamiltonian.

\subsection{Standard Mean Field Approach (MFA1)}
To implement the standard MFA for the GF model, we regard
the free energy function~(\ref{eq:finkel}) as an effective Hamiltonian.

The trial Hamiltonian  we choose, corresponds to applying 
an inhomogeneous
external field with strengths ${\bf x} = \{x_1,...,x_L\}$ along the chain  
\begin{equation}
H_0 = \sum_{i=1}^L x_i s_i \,,
\label{eq:H_0mfa1}
\end{equation} 
with $x_i$ to
be determined by minimizing the variational free-energy\cite{VarMF}
\begin{equation}
F_{var}({\bf x},T) = \sum_{i=1}^L f_0(x_i,T) + 
\langle H - H_0 \rangle_0 \,, 
\end{equation}
where $\sum_i f_0(x_i,T)$ is the free energy associated to $H_0$,
\begin{equation}
f_0(x_i,T) = -\frac{1}{\beta} \ln \bigg\{1 + \exp(-\beta x_i)\bigg\}\;.
\end{equation}  
Thermal averages, performed through the Hamiltonian $H_0$,
factorize 
$\langle s_i s_j ... s_k \rangle_0 = 
\langle s_i \rangle_0 \langle s_j \rangle_0 ... \langle s_k \rangle_0$.    
The approximate average site ``magnetization''
$m_i = \langle s_i \rangle_0$  depends only on the field $x_i$,
and is given by
\begin{equation}
m_i = \frac{\de F_0}{\de x_i} = \frac{1}{1 + \exp(\beta x_i)}\,.
\label{eq:m_i}
\end{equation}
Instead of working with external fields $x_i$'s, it is more intuitive
to use the corresponding ``magnetizations'' $m_i$'s, writing $F_{var}$
as a function of the $m_i$'s.
Due to the choice of $H_0$, Eq.~(\ref{eq:H_0mfa1}), and to the
expression~(\ref{eq:m_i}),
evaluating  the thermal average $\langle H \rangle_0$ amounts to 
replacing, in the Hamiltonian Eq.~(\ref{eq:finkel}),  
each variable $s_i$ by its thermal
average $m_i$~(\ref{eq:m_i}).
In the end we get:
\begin{equation}
F_{var}({\bf m},T) = \eps \sum_{ij} \Delta_{ij} m_i m_j -
T S({\bf m}) + R T \sum_{i=1}^L g(m_i)\,,
\label{eq:Fmvar}
\end{equation}
where $g(u) = u \ln(u) + (1-u) \ln(1-u)$ and $S({\bf m})$ is obtained
from Eq.~(\ref{eq:S}) by substituting $s_i \rightarrow m_i$. 
The last term corresponds to 
$ F_0 - \langle H_0 \rangle_0$ in Eq.~(\ref{Eq:genericFvar}):
it is the entropy
associated to the system with Hamiltonian $H_0$ and is the typical
term that stems from this kind of MFA.\cite{VarMF} 
Carrying out the minimization of function (\ref{eq:Fmvar})
with respect to ${\bf m}$ leads to 
self-consistent equations:
\begin{equation}
g'(m_i) =  \eps \sum_{j} \Delta_{ij} m_j -
R T \bigg(q - \frac{\de S_{loop}({\bf m}) }{\de m_i}\bigg)\,.
\label{eq:self}
\end{equation}
Equations~(\ref{eq:self}) can be solved numerically
by iteration and provide the
optimal values of the magnetizations that we denote by ${\bf m}^*$.
Once the set of solutions ${\bf m}^*$ is available, we can compute the 
variational free-energy  $F_{var}({\bf m}^*)$
that represents the better estimation of the system free-energy $F$.

In a mean-field approach, the  (connected) correlation function
between residues $i$ and $j$,
\begin{equation}
c_{ij}(T) = \langle s_i s_j \rangle - \langle s_i \rangle \langle s_j
\rangle \, ,
\label{eq:def_c_ij}
\end{equation}
can be recovered through a differentiation 
of $F_{var}({\bf m},T)$: 
\begin{equation}
c^{-1}_{ij}(T) = \beta
\bigg (\frac{\de F_{var}}{\de m_i \de m_j}\bigg)_{{\bf m}^*}\,,
\label{eq:standard_cij}
\end{equation}
where the subscript  indicates that the derivative is evaluated 
on the solutions ${\bf m}^*$.
Explicitating each term of $F_{var}$ we obtain the expression
\begin{equation}
c^{-1}_{ij}(T) = \frac{\delta_{ij}}{m^*_i (1 - m^*_i)} + 
 \eps \beta \Delta_{ij} -
\bigg (
\frac{\de^2  S_{loop}({\bf m}) }{\de m_i \de m_j}
\bigg )_{{\bf m}^*} \,.
\end{equation}
The correlation function matrix is given by the inversion of above matrix. 

\subsection{Second Mean Field Approach (MFA2)}
The quality of the MFA improves when we make a less naive
choice for $H_0$. One of the possible $H_0$ is suggested by the
Mu\~noz-Eaton model \cite{MEnature,MEpnas1,MEpnas2} that was proven
to be fully solvable in Ref.~\onlinecite{Brusco}. 
In fact,  even if the two models are not equivalent, 
there is  an interesting formal relationship between that model and the
present one. 
In the Mu\~noz-Eaton model, the (effective) energy of a configuration 
results from the contributions coming from the stretches of
contiguous native residues it presents, plus an entropic contribution 
from each of the non-native residues.\cite{MEpnas1,Brusco}
 
Here the effective energy Eq.~(\ref{eq:finkel}) boils down to the
contributions of stretches of contiguous non-native residues
(the loops), plus the sum of pairwise interactions of 
native residues. This latter term makes the model 
harder to solve than Mu\~noz-Eaton's one. 
If we neglect this interaction, and replace it with a residue-dependent 
contribution, the model can be mapped on the Mu\~noz-Eaton model.
Indeed, a trial Hamiltonian of the kind:
\begin{equation}
H_0({\bf x}) = \sum_{i=1}^L  x_i s_i - T S({\bf s}) \,, 
\label{Eq:H0easy}
\end{equation}
with $S(\{s_i\})$ given by Eqs.~(\ref{eq:S},\ref{Eq:Sloop}), 
can be recast as
$H_0 = C + H_{ME}$ upon the substitution $s_i \rightarrow (1-s_i)$, 
where $C=\sum x_i$ is a constant, and 
\begin{equation}
H_{ME}=\sum_{i < j} 
\left(u_{ij} \, \prod_{k=i}^{j} s_k \right) \,+\, \sum_i \mu_i s_i \,,
\end{equation}
with 
\begin{align}
u_{ij}=&- R T
[J_{i-1,j+1}-J_{i,j+1}-J_{i-1,j}+(1-\delta_{i,j-1}) J_{i,j}]\,,\\
\mu_i=& -R T(q+ J_{i-1,i+1}) -x_i\,,
\end{align}
(here $J_{i,j}=J(r_{ij})$ of Eq.~(\ref{Eq:J}); $J_{0,i}=J_{i,L+1}=0$).
Now the trial Hamiltonians reads formally as 
the Mu\~noz-Eaton Hamiltonian: see Eq.~(1) of Ref.~\onlinecite{Brusco},
where the symbol $m_i$ was used instead of $s_i$.

Hence, we choose Eq.~(\ref{Eq:H0easy}) as the trial Hamiltonian, and 
write down the mean field equations Eq.~(\ref{Eq:meanfield}):
\begin{equation}
\tilde{\varepsilon_l} \left[ \sum_{i<j} \varepsilon_{i,j}
\Delta_{i,j}(C_{i,j,l}-C_l C_{i,j}) - 
\sum_{i=1}^L x_i (C_{i,l}-C_i C_l) \right] =0
\label{Eq:mfeasy}
\end{equation}
for $l=1, \ldots, L$.
These equations involve the functions 
\begin{subequations}
\begin{gather}
 C_{i} =  \langle s_i\rangle_0\\
 C_{i,j} = \langle s_i s_j \rangle_0\\ 
 C_{i,j,l} = \langle s_i s_j s_l \rangle_0
\end{gather}
\label{Eq:Ceasy}
\end{subequations}
where averages are evaluated by the same transfer matrix
technique as in Ref.~\onlinecite{Brusco}.

Using the fact that CVM is exact for the Mu\~noz-Eaton Model, 
it can also be proven that the three-point functions $ C_{i,j,l}$ 
can be written as a function of the two-point ones: $C_{i,j,l} = C_{i,j}
C_{j,l}/C_j$, for $i<j<l$~\cite{Aleobs}. 
This greatly reduces the computational cost of minimizing the variational 
free energy and makes the approach particularly suitable for  
long polypeptide chains.

Correlations $c_{ij}$ could still be evaluated as in
Eq.~(\ref{eq:standard_cij}), but now the dependence of
$F_{var}$ upon $m_i$ cannot be worked out explicitly, and the
derivatives must be evaluated resorting to the dependence on the
fields $x_j$: namely $\partial F_{var}/ \partial  m_i = \sum_j
(\partial x_j/\partial m_i)( \partial  F_{var}/\partial x_j)$.
However, this entails to evaluate the four-point averages 
$ \langle s_i s_j s_k s_l \rangle_0$, with a consequent 
relevant computational cost, for this reason, we will not pursue this 
strategy in the following.  

\subsection{Third Mean Field Approach (MFA3)}
In the previous MFA version, the entropic term was treated exactly
while the energy contribution was very roughly approximated. 
This new version aims to better incorporate the energy contributions
and we shall see that results are in excellent agreement with the
exact solution obtained by exact enumeration on the $\beta$-hairpin.
We consider the set of configurations of the proteins with $M$
native residues  ($M=0,...,L$). 
We then take as the trial Hamiltonian
\begin{equation}
H_0({\bf x}) = \sum_{M=0}^L \delta(M - \Sigma_{i} s_i)
H_0^{(M)}({\bf x}) \,,
\label{Eq:H0difficult}
\end{equation}
where $\delta(\bullet)$ is the Kronecker delta,  and
$H_0^{(M)}$ is the Hamiltonian restricted  to the configurations
with $M$ natives: 
\begin{equation}
H_0^{(M)}({\bf x})= \sum_{i=1}^L \tilde{\varepsilon_i}\, x_i
\frac{M-1}{L-1} s_i - T S({\bf s}) \,, 
\label{Eq:H0M}
\end{equation}
with   $\tilde{\varepsilon_i} = (1/2) \sum_{j=1}^N \varepsilon_{i,j}
\Delta_{i,j}$.
Each  residue $i$, in a generic configuration with
$M$ native residues, feels an interaction $\tilde{\varepsilon_i}$
which it would feel in the native state, weakened by a factor $(M-1)/(L-1)$
(accounting for the fact that not all the residues are native),
times  the external field $x_i$, to be fixed by the mean field procedure.

This scheme is useful for taking correlations into account 
in a better way than in the usual MFA, so to gain some insight on the 
parts of the chain that fold first and to investigate folding pathways.
In this framework the partition function is:
\begin{equation}
{\cal Z}_0 = \sum_{M=0}^L Z^{(M)} = \sum_{M=0}^L
\sum_{\{s_i=0,1\}}^{\;\;\;\;\;\;\;\;\;(M)} 
\exp(- \beta H_0^{(M)})\,,
\label{Eq:Zdifficult}
\end{equation}
where the symbol $(M)$ above the sum  indicates that
the sum is restricted to configurations with $M$ native residues.
The mean field equations (\ref{Eq:meanfield}) reads
\begin{equation}
\tilde{\varepsilon_l} \left[ \sum_{i<j} \varepsilon_{i,j}
\Delta_{i,j} ({\cal
C}'_{i,j,l} - {\cal C}_{i,j} {\cal C}'_l) - 
\sum_{i=1}^L x_i \tilde{\varepsilon_i} ( {\cal C}''_{i,l} - {\cal C}'_i
{\cal C}'_l)\right]=0 \, , 
\label{Eq:mfdifficult}
\end{equation}
for each $l$, where 
\begin{subequations}
\begin{gather}
{\cal C}_{\bullet}=\sum_{M=1}^L C_{\bullet}^{(M)}\\
{\cal C}'_{\bullet}=\sum_{M=1}^L \frac{(M-1)}{(L-1)} C_{\bullet}^{(M)} \\
{\cal C}''_{\bullet}=\sum_{M=1}^L \frac{(M-1)^2}{(L-1)^2} C_{\bullet}^{(M)} \\
 C_{i}^{(M)} = \frac{1}{{\cal Z}_0}
 \sum_{\{s_i\}}^{\;\;\;\;\;\;\;\;\;(M)} s_i \exp(-\beta H_0^{(M)})\\
 C_{i,j}^{(M)} = \frac{1}{{\cal Z}_0}
 \sum_{\{s_i\}}^{\;\;\;\;\;\;\;\;\;(M)} s_i s_j 
\exp(-\beta H_0^{(M)})\\
 C_{i,j,l}^{(M)} = \frac{1}{{\cal Z}_0}
 \sum_{\{s_i\}}^{\;\;\;\;\;\;\;\;\;(M)} s_i s_j s_l \exp(-\beta
 H_0^{(M)})
\end{gather}
\label{Eq:Cdifficult}
\end{subequations}
are the contributions to the correlation associated to configurations 
with $M$ native residues. 
The transfer-matrix method applied in Ref.~\onlinecite{Brusco} allows keeping
track separately of the contributions 
coming from the configurations with a given total number of native residues,
therefore it is possible to evaluate exactly the partition
functions ${\cal Z}_0^{(M)}$, and all the averages
Eq.~(\ref{Eq:Cdifficult}) involved in the mean field equations
Eq.~(\ref{Eq:mfdifficult}). The
computational cost is relevant, though: in fact, due to the necessity
of evaluating all $C_{i,j}^{(M)}$ and some $C_{i,j,l}^{(M)}$ (the ones
actually occurring in Eq.~(\ref{Eq:mfdifficult})), $O(L^6)$ elementary
multiplications are required.  
As far as correlations $c_{ij}$ are concerned, the same discussion of
the MFA2 case holds.

\section{The $\beta$-Hairpin}
We compare the MFA results with numerical simulations on
the $\beta$-hairpin, the fragment 41$-$56 of the naturally
occurring protein GB1 (2GB1 in the  Protein Data Bank).\cite{seqhair}
This peptide has been widely studied experimentally,\cite{MEnature,
Honda2000,Koba2000}
through all-atom simulations\cite{Karplus,Klimov,Pande}
and simplified models.\cite{MEnature,MEpnas1,Levine} Thus 
it represents a good test for the validity of the model and its 
approximations.
Since the $\beta$-hairpin contains only $L=16$ aminoacids, 
we can carry out exact enumeration over the $2^{16}=65536$ possible 
configurations to compute explicitly the partition function 
$$
Z(\beta) = \sum_{\{s_i\}} \exp(-\beta H)
$$
of the model. Once the function $Z$ is known,
all the thermal properties are available and it is possible to
completely characterize the thermal folding of the hairpin peptide. 
However, first, we have to adjust the model free parameters
$\varepsilon$ and $q$ to reproduce experimental data on the
hairpin equilibrium folding. Experimental results
on tryptophan fluorescence,\cite{MEpnas1} show that,
in the folded state, the 99\% of molecules contain a well formed
hydrophobic cluster made of Trp$43$, Tyr$45$, Phe$52$ and Val$54$.
In the model, the formation of the hydrophobic cluster
is described by the behaviour of the four-points 
correlation function $Q_{hyd} = \langle s_3 s_5 s_{12} s_{14}\rangle$ 
(notice that, here and in the following, 
residues are renumbered from 1 to 16, instead of 41$-$56).
\begin{figure}
\includegraphics[clip=true,width=\columnwidth, keepaspectratio]
{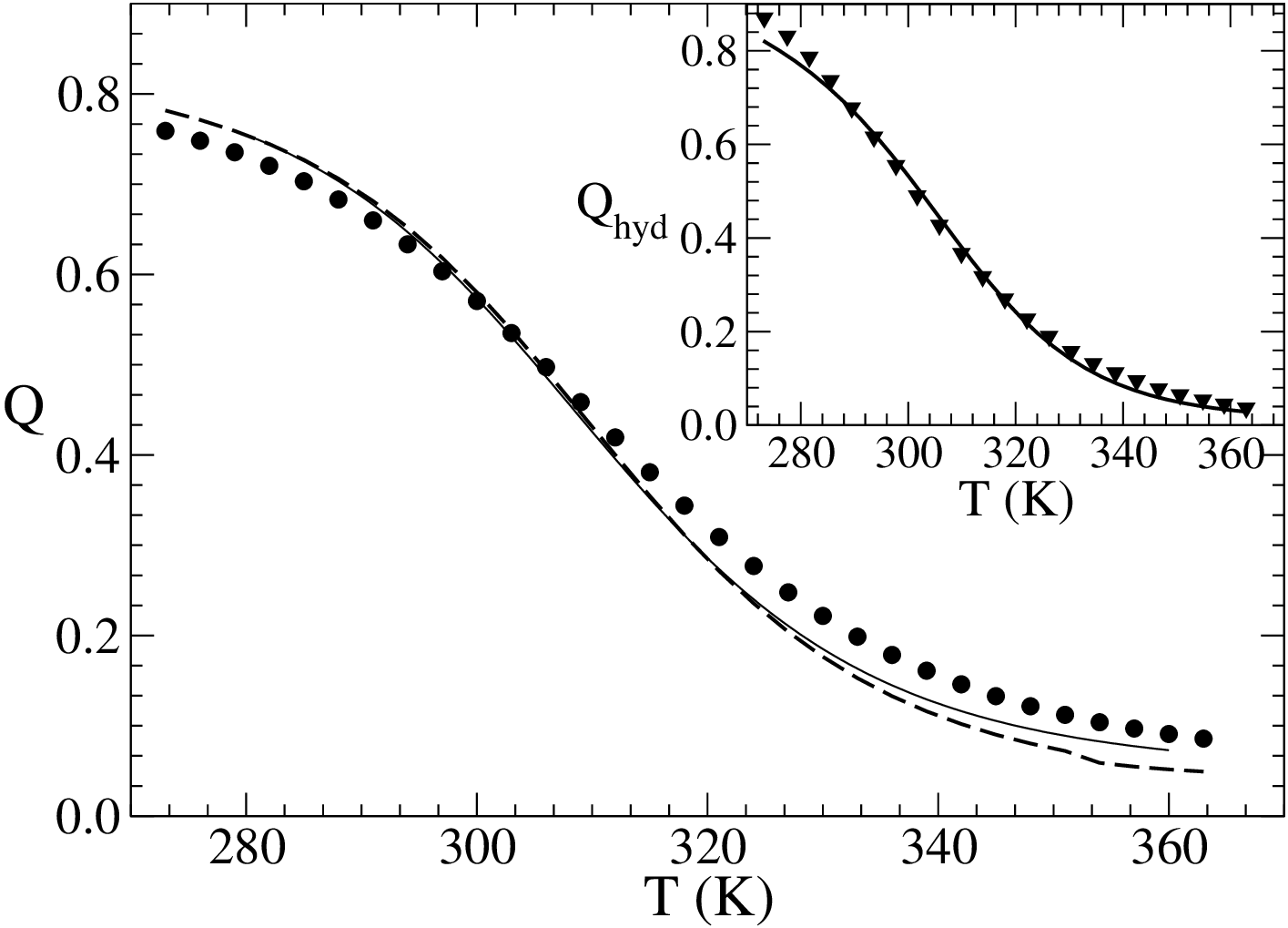}
\caption{
\label{fig:order}
Fraction of native residues $Q$ (see \ref{eq:Q}) during thermal folding,
according to the GF model. Full dots are the exact result 
obtained by exhaustive enumeration. Dashes and full lines indicate 
 MFA1 and  MFA3 approximations,
respectively.
Inset: Fit of the hydrophobic cluster ($W43-Y45-F52-V54$) population
$Q_{hyd} $ (solid) to the experimental data from~\cite{MEnature} 
(triangles).
}
\end{figure}
The choice of the model parameters $q=2.32$ and
$\varepsilon=-0.0632$ (kcal/mol) provides the best fit of
$Q_{hyd}$ to the behavior of the experimental fraction of folded molecules 
(cfr. inset of Fig.~\ref{fig:order} with Fig.~3 of Ref.~\onlinecite{MEpnas1}). 
We can now assess the goodness of the model and its  mean-field
approximations, by comparing their predictions with the experimental
results and simulations. 

Averages and correlations 
within the mean-field schemes will be evaluated as follows:
for MFA1, the self-consistent mean-field equations~(\ref{eq:self}) 
are solved by
iteration, substituting an arbitrary initial value for  ${\bf m}$ at
the right-hand  side of Eq.~(\ref{eq:self}), evaluating $m_i$ from the
left-hand side, and 
substituting again the latter value in the right-hand side, until
convergence is achieved.
 
\begin{figure}
\includegraphics[clip=true,width=\columnwidth, keepaspectratio]
{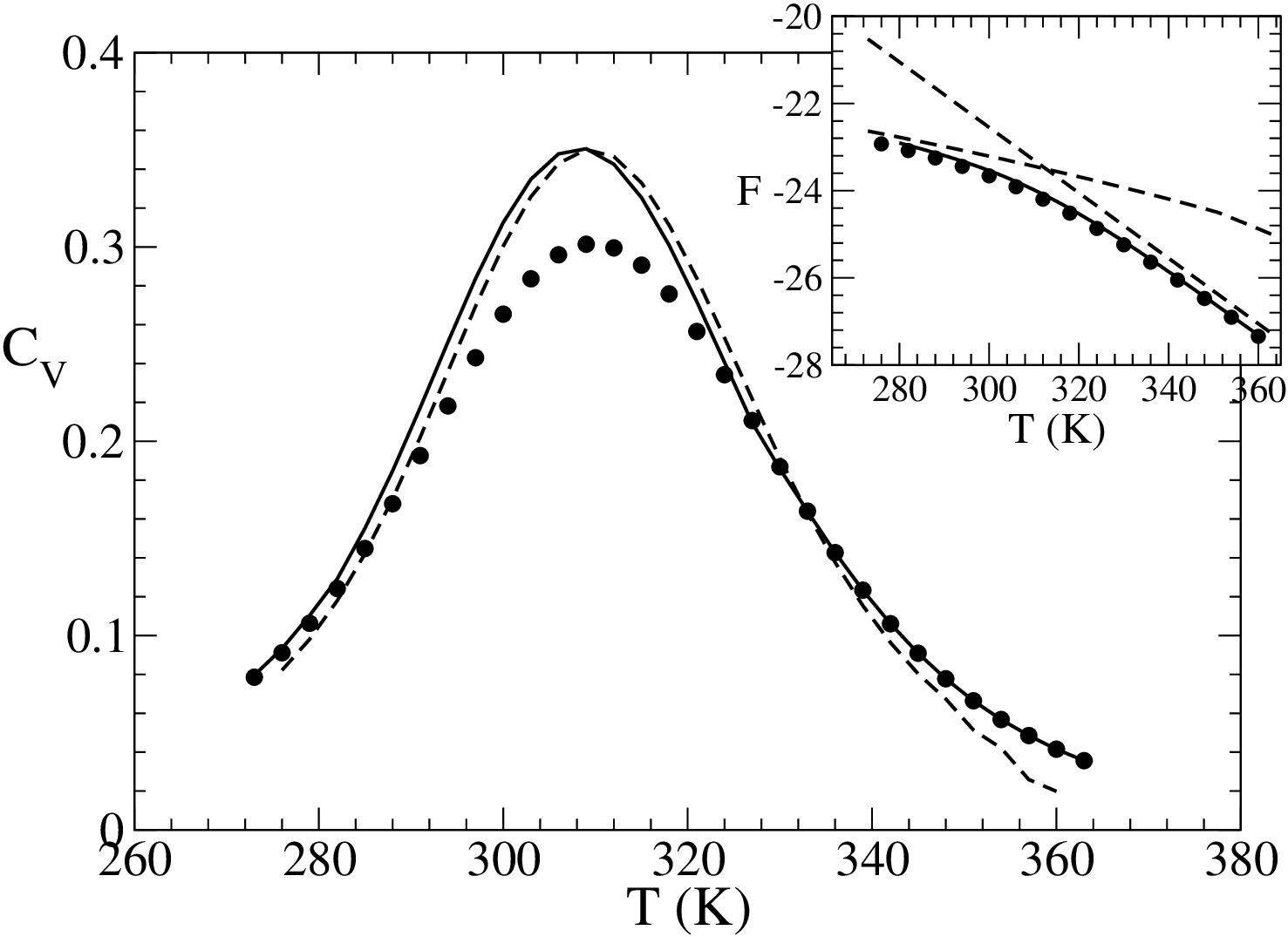}
\caption{
\label{fig:cfr}
Comparison between MFA and the exact enumeration.
Behaviour of specific heat (kcal~mol$^{-1}$~K$^{-1}$) and 
free-energy (kcal mol$^{-1}$) with temperature, as obtained
by the exact enumeration of the GF model applied to the 
hairpin. Dots indicate the exact results, while dashed and solid
lines correspond to MFA1 and MFA3, respectively.
In the inset, the mean-field free energies and the exact 
free energy are plotted against temperature rendering conventions
as before. Notice the crossing of two branches of MFA1 at the 
transition temperature. 
}
\end{figure}

In the present case, this procedure converges
quickly to two different solutions (depending on the starting values
of the fields), 
corresponding to different phases: the 
folded one ($m_i \sim 1$) at low temperature and the unfolded ($m_i \sim 0$) 
at high temperature. 
Starting from the unfolded phase and lowering the temperature the 
solution of Eqs.~(\ref{eq:self}) remains trapped into a set of 
misfolded metastable states.
Only at temperatures well below the folding temperature $T_F$ the
solution collapses into the one representing the folded state.
The opposite happens when the temperature is increased starting 
from the folded phase.
This is a typical scenario of first-order like transitions, which is 
reproduced by the mean field approach.
The situation is well illustrated by the behaviour of the mean field 
free-energy, which exhibits two branches $F_1(T)$ and $F_2(T)$ as 
shown by the dashed lines in the inset of Fig.~\ref{fig:cfr}. 
The intersection of the two branches defines the
mean-field folding temperature. 
At a given temperature, the free-energy of the protein is obtained
by selecting the minimum of the two branches 
\begin{equation}
F(T) = \min \{F_1(T),F_2(T)\}.
\label{eq:fren}
\end{equation}

In this approximation other observables present a jump at transition:
this reflects the fact that in the thermodynamic limit (here
corresponding to  infinitely long proteins), only the solution with 
the lowest free-energy would be
physical. To take into account finite-size effect, 
we decide to introduce an interpolating formula to 
deal with a continuous quantity:   
\begin{equation}
\langle O \rangle = \frac{\e^{-\beta F_1} \langle O \rangle_1 +
\e^{-\beta F_2} \langle O \rangle_2}
{\e^{-\beta F_1} + \e^{-\beta F_2}}\,, 
\label{eq:matching}
\end{equation}
where $\langle O \rangle_1$ and $\langle O \rangle_2$ are the averages of 
the observable in the above mentioned branches. 
In this way we compute the 
average magnetization (i.e. the fraction of correctly folded residues) 
of the protein:
\begin{equation}
Q = \frac{1}{L} \sum_{i=1}^L \langle s_i\rangle\,, 
\label{eq:Q}
\end{equation} 
as well as its 
energy $\langle E \rangle$.  In the latter case  $\langle E
\rangle_1$, $\langle E
\rangle_2$  are evaluated as  $\langle E
\rangle_\alpha = \partial (\beta F_\alpha) / \partial \beta$.

Differentiating the energy with respect to the temperature, 
we get the specific heat, reported in Fig.~\ref{fig:cfr}.
Notice that this is the correct recipe to take into
account also the contributions to the specific heat coming from the
change of the native fraction of molecules: the alternative one,
obtained with the direct application of  
Eq.~(\ref{eq:matching}) to the specific heats
$C_v^1=\partial \langle E \rangle_1 / \partial T$ and $C_v^2=\partial
\langle E \rangle_2 / \partial T$, would neglect the change in the
number of folded molecules, and account only for the variations of the
energy within the pure native or unfolded state.  
For the same reason, Eq.~(\ref{eq:matching}) is not useful to
match the correlation functions $c_{ij}$ evaluated on the two
branches. It would yield only a linear superposition of the $c_{ij}$'s 
relative to native and unfolded states, while the correct functions 
should account for the contributions coming from 
all the configuration space.

Coming to MFA2, we observe that it keeps exactly into account the
entropic term Eq.~(\ref{eq:S}). Yet, solving the mean-field equations
yields again two different solutions at each temperature. Thus, MFA2
presents the same kind of problems in characterizing the folding transition
states as MFA1. This is why in the following we will present results
just for MFA1 and MFA3, that behave in a substantially different way.

With MFA3, in fact, a unique set of fields ${\bf x}(T)$ is observed,
independent of the starting values,  for
any temperature in the interesting range around the
transition, and no empirical connection rule
Eq.~(\ref{eq:matching}) is required.  
Moreover, at odds with  MFA1 and MFA2, the difference between $F_{var}$
and $F_0$ in Eq.~(\ref{Eq:genericFvar}) happens to be negligible at
all the relevant temperatures: 
$F_0$ is a very
good approximation to $F_{var}$. This suggests that the correct correlation
functions, which would be very hard to evaluate, can be replaced by
the ones  involving averages with the trial Hamiltonian $H_0$: $c_{ij}
\simeq \langle s_i s_j \rangle_0 -\langle s_i \rangle_0 \langle s_j
\rangle_0 $. Thus, within MFA3 it is possible to give a substantially
correct characterization both of the native and unfolded states, and
of the folding nucleus. 

In Fig.~\ref{fig:order} we plot $Q$ of Eq.~(\ref{eq:Q}) as a function
of the temperature, for the original model, for MFA1 (with the help of
Eq.~(\ref{eq:matching})) and MFA3.   
At low temperatures, where the protein assumes its native
state, $Q = 1$,  while $Q\sim 0$ in coil configurations
(i.e. at high temperatures).
Mean field approximations appear to be slightly more ``cooperative''
than the original model, according to their steeper 
sigmoidal shape. The temperature at which
$Q = 1/2$ is an estimate of the folding temperature: we have 
$T_F\sim 306-306.5$ K for the original model, and  $T_F\sim 305$ K 
for both MFA1 and MFA3. 

In Fig.~\ref{fig:cfr} we plot the specific heat: 
\begin{equation}
C_v = \frac{\langle E^2 \rangle - \langle E \rangle^2}{RT^2} =
\frac{\partial U}{\partial T}
\end{equation}
and the free energy.
The peak of $C_v$, which provides another definition for the folding 
temperature, occurs around  $T_F \sim 309.5$ K for the exact
model and its mean field approximations. 
Notice that MFA1 and MFA3 substantially recover the position of the
exact peak, even if the transition appear a little sharper in the
mean-field cases.

The above estimates of the folding temperatures are somewhat higher
than the experimental ones, $T_F \sim 298 $ K
~\cite{MEnature} and $T_F \sim 295.3 $ K ~\cite{Honda2000}. 
Interestingly, $T_F$ appears to be higher than the experimental value
also for ``united atom'' simulation\cite{Klimov} ($T=308$ K in the 
Go-model case, $T=333$ K with the full potential introduced in that paper),
and for all-atoms simulations.\cite{Karplus} 

\begin{figure}
\includegraphics[clip=true,width=\columnwidth,keepaspectratio]
{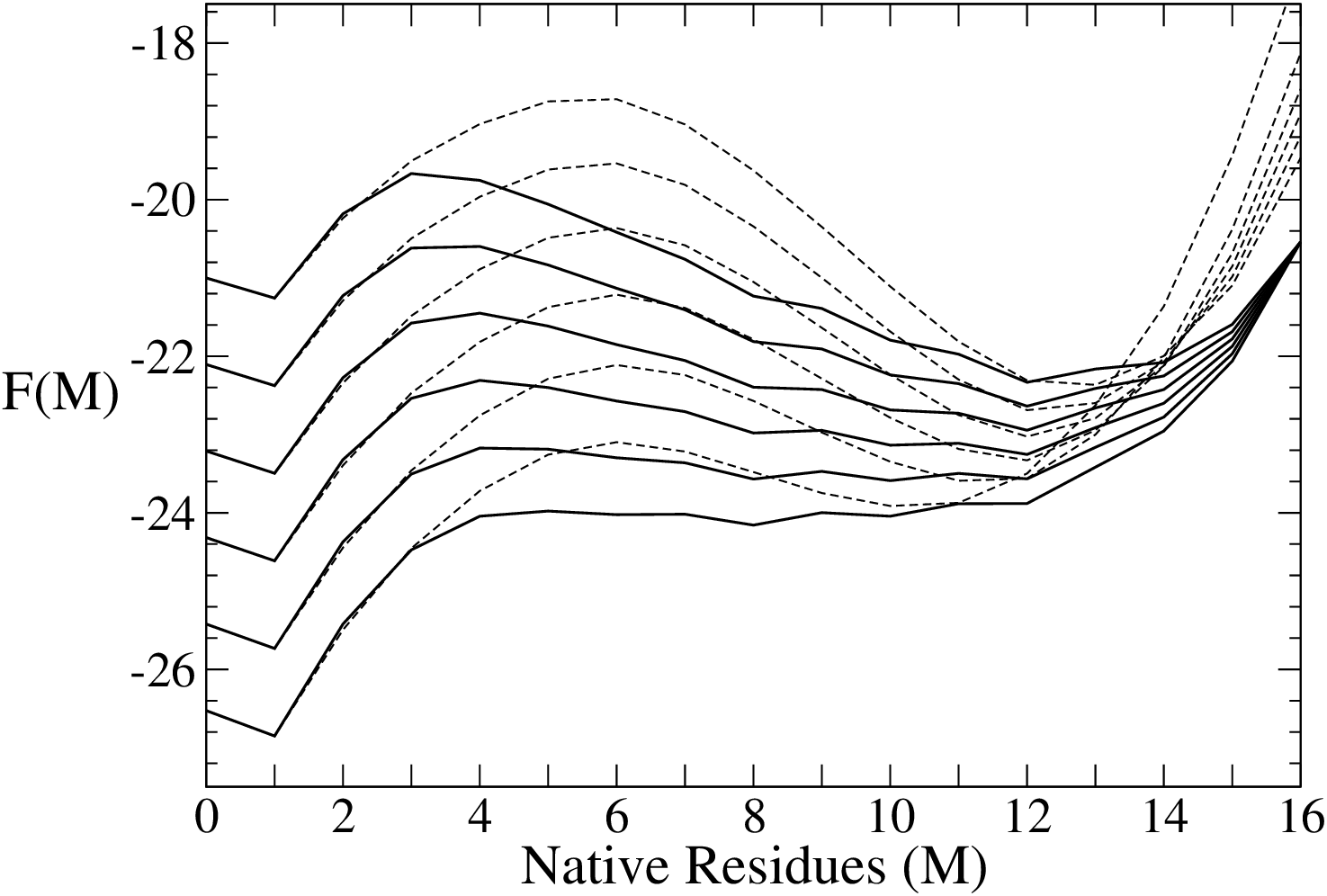}
\caption{
\label{fig:profiles}
Free energy landscape for the hairpin, 
i.e. plot the free-energy (kcal mol$^{-1}$)
of the system vs. the number of native residues $M$. 
Solid lines: exact results for the GF model; dotted lines:
MFA3. Temperatures are 285 K, 300 K, 315 K, 330 K, 345 K, 360 K, 
from top to bottom.
}
\end{figure}

Free-energy profiles, for various temperatures, are plotted in 
Fig.~\ref{fig:profiles} 
versus the number of native residues $M$, that we use as the folding reaction 
coordinate.
Profiles suggest that the $\beta$-hairpin folding is well described
by a two state process, i.e. $F(M)$ exhibits two minima separated by a
barrier that has to be overcome in order to reach the native/unfolded state.
Notice, though, that this doesn't rule out the possibility that
folding might not be a two-state process in this case: this could
happen if the number of native residues $M$ was not a good reaction
coordinate.\cite{karpnote}  
Other alternative order parameters should be considered,
in addition to $M$, to completely ascertain the nature of the transition.  

The comparison between exact and mean-field results reveals that 
the barrier appears to be overestimated in the MF scheme, where
it is also shifted towards higher values of the reaction coordinate: again,
the MFA appears to be more cooperative than the
original model. Notice however that the free-energy and position of the
native and unfolded minima, and hence the stability gap, 
are correctly recovered, especially at
temperatures close to transition (i.e. the second and third plots from
top down).

Another interesting characterization of the folding pathway comes from
the temperature behavior of the pairwise 
correlation functions between residues 
\begin{equation}
c_{ij}(T) = \langle s_i s_j \rangle - \langle s_i \rangle \langle s_j
\rangle \,, 
\end{equation}
that provides an insight on the probability of contact formation
during the thermal folding, as shown in Refs.~\onlinecite{Cecco,MCFM}

\begin{figure}
\includegraphics[clip=true,width=\columnwidth, keepaspectratio]
{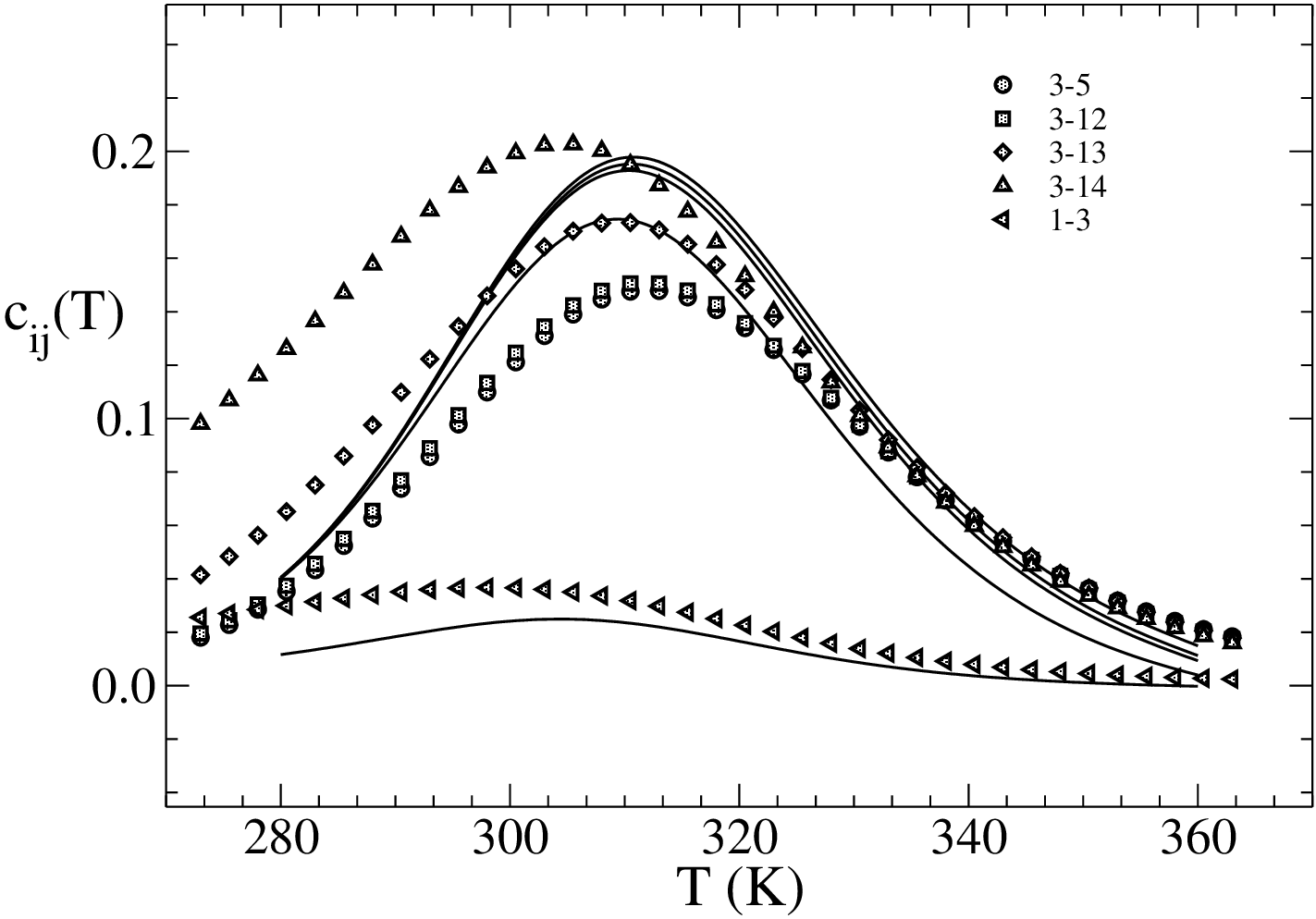}
\caption{
\label{fig:correla}
Temperature behavior of the correlation function $c_{ij}(T)$ of native 
contacts involving the Tryptophan (Trp45). Symbols correspond to 
the exact solution, while solid lines indicate the MFA3 results for the 
contacts 1$-$3, 3$-$5, 3$-$12, 3$-$13, 3$-$14, from  bottom to top.
}
\end{figure}

In fact, each function $c_{ij}(T)$ develops a peak at a characteristic 
temperature, which can be regarded as the temperature of formation/disruption 
of the contact $i-j$. In Fig.~\ref{fig:correla}, we plot the correlation
functions between Trp45 and residues to which it is in native
interaction. 
The height of each peak indicates the relevance of the contact
from a thermodynamical point of view.\cite{Cecco,MCFM,Dokholyan,Settanni}
Thus, each  contact turns out to  be characterized thermodynamically 
by the location (temperature) and the height of the corresponding
peak. This provides a criterion for ranking contacts in order of 
temperature and relevance (see Refs.~\onlinecite{Cecco,MCFM}). 
For example, at the folding temperature $T_F$, the contacts that mainly 
contribute to the folding transition must be searched among those with the 
characteristic temperature located around $T_F$ and with highest 
peak of $c_{ij}$.   
Correlation analysis for the hairpin is summarized in 
Table~\ref{tab:correla}, 
where we report the temperature and the height of correlation function 
peaks, between residues which share a native contact. 
Contacts are sorted in temperature and whenever a tie
occurs the sorting runs over the heights of the peaks.
In this way, we can have a picture of how contacts are established during
the thermal folding. 
\begin{table}
\begin{ruledtabular}
\begin{tabular}{|rll||rll|}
\multicolumn{3}{|c||}{Exact} &  
\multicolumn{3}{c|}{Mean Field} \\
Contact & T$_{char}$ & Corr. Peak  & Contact & T$_{char}$ & 
Corr. Peak \\
\hline\hline 
6-11  & 316.5 &  0.21457 &      6-11  & 312.0 & 0.21582 \\
6-9   & 316.5 &  0.21346 &      9-11  & 312.0 & 0.21553 \\
9-11  & 316.5 &  0.21243 &      6-9   & 312.0 & 0.21441 \\
6-8   & 316.5 &  0.18262 &      6-12  & 311.5 & 0.21535 \\
5-11  & 315.5 &  0.20457 &      4-6   & 311.5 & 0.21449 \\
6-12  & 315.5 &  0.20214 &      5-11  & 311.5 & 0.21449 \\
7-9   & 315.5 &  0.18462 &      11-13 & 311.5 & 0.21447 \\
6-10  & 315.5 &  0.16488 &      5-12  & 311.5 & 0.21414 \\
5-12  & 315.0 &  0.21670 &      4-12  & 311.5 & 0.21031 \\
5-7   & 315.0 &  0.18461 &      5-13  & 311.5 & 0.21007 \\
10-12 & 315.0 &  0.15697 &      5-7   & 311.5 & 0.19342 \\
7-10  & 314.5 &  0.15048 &      7-9   & 311.5 & 0.19198 \\
8-10  & 314.5 &  0.14339 &      6-8   & 311.5 & 0.18361 \\
11-13 & 314.0 &  0.17713 &      10-12 & 311.5 & 0.17117 \\
4-6   & 314.0 &  0.17575 &      6-10  & 311.5 & 0.16936 \\
4-12  & 313.5 &  0.19244 &      4-13  & 311.0 & 0.20685 \\
5-13  & 313.5 &  0.18681 &      7-10  & 311.0 & 0.15091 \\
3-5   & 312.0 &  0.14805 &      8-10  & 311.0 & 0.14385 \\
4-13  & 310.5 &  0.20932 &      3-5   & 310.5 & 0.19785 \\
3-12  & 310.5 &  0.15090 &      3-12  & 310.5 & 0.19526 \\
12-14 & 310.5 &  0.14416 &      3-13  & 310.5 & 0.19284 \\
3-13  & 309.5 &  0.17364 &      12-14 & 310.5 & 0.19111 \\
4-14  & 309.0 &  0.16487 &      4-14  & 310.0 & 0.18529 \\
2-4   & 307.0 &  0.08024 &      3-14  & 309.5 & 0.17464 \\
2-13  & 306.5 &  0.08037 &      1-15  & 308.5 & 0.08026 \\
13-15 & 306.0 &  0.05241 &      2-4   & 307.5 & 0.08349 \\
3-14  & 304.5 &  0.20284 &      2-13  & 307.0 & 0.08081 \\
2-14  & 300.0 &  0.11045 &      2-14  & 306.0 & 0.07482 \\
1-3   & 298.5 &  0.03678 &      13-15 & 306.0 & 0.04143 \\
1-14  & 296.5 &  0.03570 &      2-15  & 305.0 & 0.01189 \\
14-16 & 290.5 &  0.09773 &      1-3   & 304.5 & 0.02493 \\
2-15  & 273.0 &  0.12746 &      1-14  & 304.5 & 0.01934 \\
1-15  & 273.0 &  0.05005 &      14-16 & 291.0 & 0.00293 \\
\end{tabular}                 
\end{ruledtabular}
\caption{\label{tab:correla}
Ranking of native contacts according to
characteristic temperature and height of the correlation
peak.\cite{MCFM} 
Contacts 1-16
and 2-16 have been neglected: they yield bad results because they are not 
stable even in the experimental native structure.\cite{MEpnas1}
The first three columns refer to the exact solutions, the others to 
MFA3 results.
}
\end{table}
Assuming that the order of contact stabilization upon
decreasing the temperature reflects the order of formation during
folding, this is also a qualitative indication of the folding pathway.

We see, from the first three columns of Table~\ref{tab:correla}, that
GF model predicts that the $\beta$-hairpin folding begins with the 
formation of contacts 6$-$11 and 6$-$9, 9$-$11
and 6$-$8, located in the region between the turn (8$-$9) and the
hydrophobic cluster. Then, upon lowering the temperature, 
the folding proceeds with the formation of the other contacts that 
complete $\beta$-hairpin structure. 
This is at odds with the results of more detailed models and 
simulations~\cite{Karplus,Klimov,Pande} predicting that folding starts 
with the formation of contacts between the side chains of
the hydrophobic cluster, and proceeds with the
stabilization of the hydrogen bonds in the loop region (there is no
agreement on the order of hydrogen-bonds formation, though).
GF model predictions are different also from those of the Mu\~noz-Eaton 
model,\cite{MEpnas1,Brusco} where the hairpin starts
folding from the loop region and proceeds outwards in a zipper fashion.
Experimental results relying on point mutations\cite{Koba2000} 
witness the importance of the hydrophobic residues 3, 5, 12 and, 
to a minor extent, 14, in stabilizing the hairpin structure. 
Remarkably, contacts between residues 6, 9, 11 appear to be partially 
present also in denaturing conditions.\cite{Koba2000}   

It is interesting to notice, however,  that, according to Table 
\ref{tab:correla}, contacts 3$-$13, 4$-$14, 3$-$12, 12$-$14, 4$-$13 of 
the hydrophobic cluster are mainly 
established  around the folding temperature, which
suggests that also in GF model the hydrophobic cluster plays a
central role.
This is a nice feature of the model because
it is consistent with the experimental evidence (fluorence signal)
for the formation of the tryptophan hydrophobic environment at the 
folding.

The estimation of correlation functions provided by MFA3 is only 
in qualitative agreement with exact results (see 
Fig.~\ref{fig:correla}): contacts are formed in a narrower range of
temperatures, and a direct comparison would be meaningless. 
However we can ask what kind of information can be extracted 
from the mean-field results, wondering, for instance, whether the ranking 
of contact formation provided by MFA3 is ``statistically equivalent'' to 
that given by exact solution. 
Thus, we apply the Spearman rank-order correlation test.\cite{NumRec} 
This test amounts to computing Spearman correlation 
\begin{equation}
R_s =  1 - \frac{6 \sum_{i=1}^n (x_i-y_i)^2}{n(n^2-1)} \,,
\end{equation}
where $x_i$, and $y_i$ are the integer indicating the positions of 
the $i$-th contact in the 
two ranking respectively. The parameter, $R_s$ is $1$ when the order in 
the two ranks is the same 
$x_i=y_i$, while $R_s = -1$, when the order is reverse $x_i+y_i=n$.
For data in Table~\ref{tab:correla}, we obtain the value $R_s = 0.902$, 
that has a probability $P < 10^{-6}$ to take place if the null hypothesis 
of uncorrelated ranks holds.
This indicates that the order between the contacts obtained with exact and
approximate methods is extremely significative: the mean-field
approach basically recovers the correct order of contact formation and
relevance as obtained with the true original model. 

One of the most important experimental techniques for characterizing the
folding nucleus of a protein (more precisely of a protein with two-state 
folding) consists in the evaluation of $\Phi-$values. 
$\Phi-$values measure the effect of "perturbation" introduced in 
a protein by site-directed mutagenesis.\cite{AF-Book}    
A mutation performed on the $i$-th residue may affect 
the thermodynamics and kinetics, by altering the free-energy
difference between the native and unfolded state (i.e. the stability
gap) or the height of the folding/unfolding barrier.
Its effect is quantified through the $\Phi$-value, defined as    
\begin{equation}
\Phi_i(T) =
\frac{\Delta(\Delta F_{\ddag U})}{\Delta(\Delta F_{N U})}\;,
\label{eq:phi}
\end{equation} 
where  $\Delta F_{\ddag U}= F_\ddag -F_U$, $\Delta F_{N U}= F_N
-F_U$, and 
$\Delta(\Delta F_{\ddag U})$ and $\Delta(\Delta F_{N U})$ are the
variations, with respect to the wild type protein, introduced by the
mutation in the folding barrier and stability gap.
Experimentally, $\Delta(\Delta F_{\ddag U})$  is derived from the
changes in the kinetic rates induced by different denaturant
concentrations, while $\Delta(\Delta F_{N U})$ is extracted from the 
changes in the equilibrium population.
$\Phi$-values are different for different mutations of
a residue; in any case, a $\Phi$-value close to one implies that the 
mutated residue has a native-like environment in the transition state 
and hence is involved in the folding nucleus. A value close to 
zero, instead, indicates that the transition state remains unaffected by the 
mutation, and hence the mutated residue is still unfolded at transition.

\begin{figure}
\includegraphics[clip=true,width=\columnwidth,keepaspectratio]
{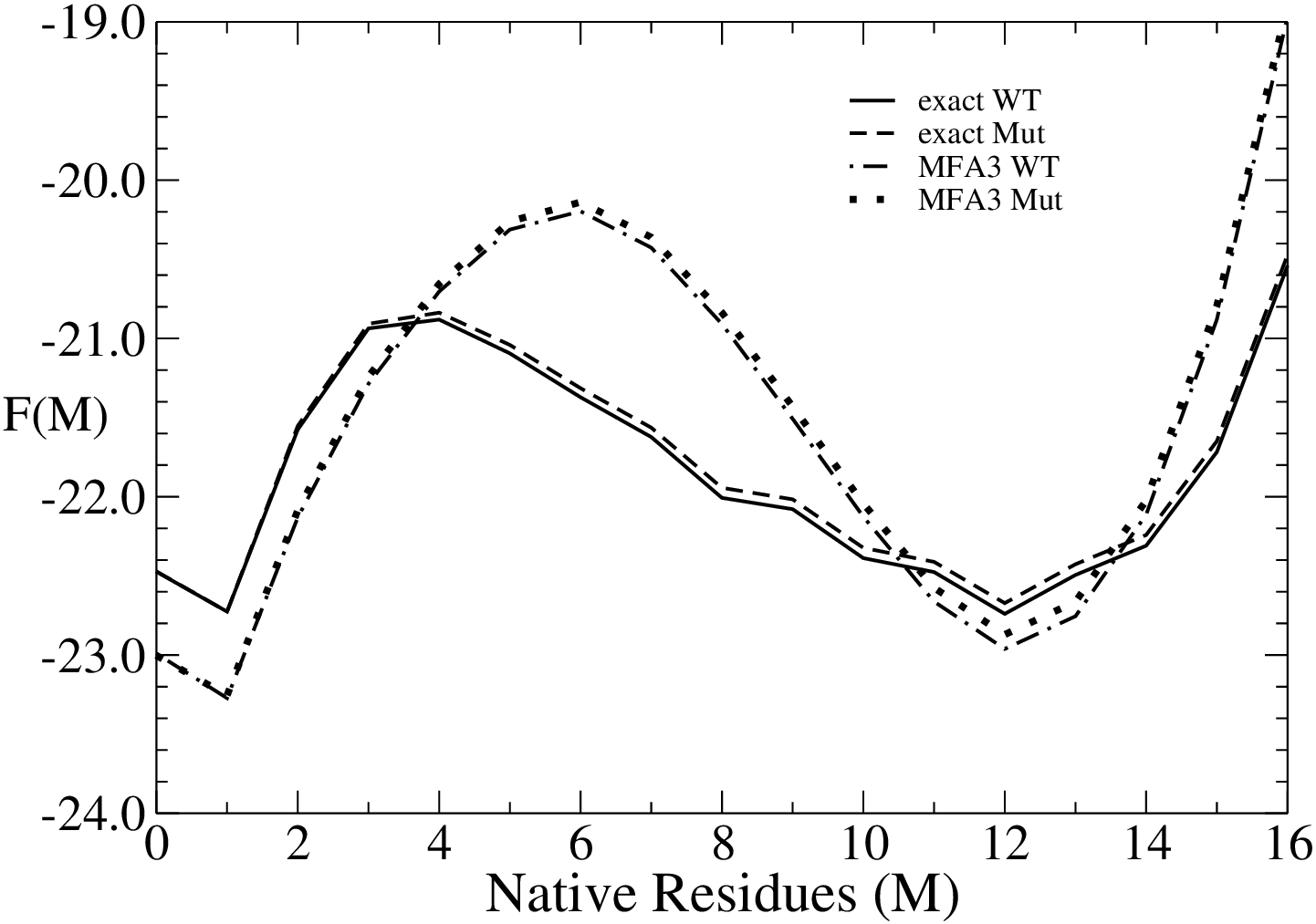}
\caption{
\label{fig:varprof} 
Variation on the free-energy profile 
induced by the perturbation on all the interactions involving the 
sixth residue (Asp46) of the Hairpin. The variation (in kcal mol$^{-1}$) 
is computed for both the exact solution and MFA3, at the respective
temperatures of equal populations of the native and unfolded basins. 
Solid and dashed lines indicate wild-type and mutated profiles 
respectively for the exact solution; dot-dashed and points 
refer to wild-type and mutated profile 
respectively in the MFA3.
}
\end{figure}

In our theoretical description, a mutation at site $i$ is simulated
by weakening the strength of the couplings $\varepsilon \Delta_{ij}$ of
$1$\% between residue $i$ and the others.
We choose a small perturbation because we cannot predict what kind
of rearrangements in the local structure, and hence in the contact map,
a true residue-to-residue mutation would involve.
Our choice warrants that the effect of mutation remains local and does not 
disrupt completely the state. 
In figure~\ref{fig:varprof}, we show
the effect of a ``mutation'' of the sixth residue (Asp46) on the
free-energy profiles. 

\begin{figure}
\includegraphics[clip=true,width=\columnwidth,keepaspectratio]
{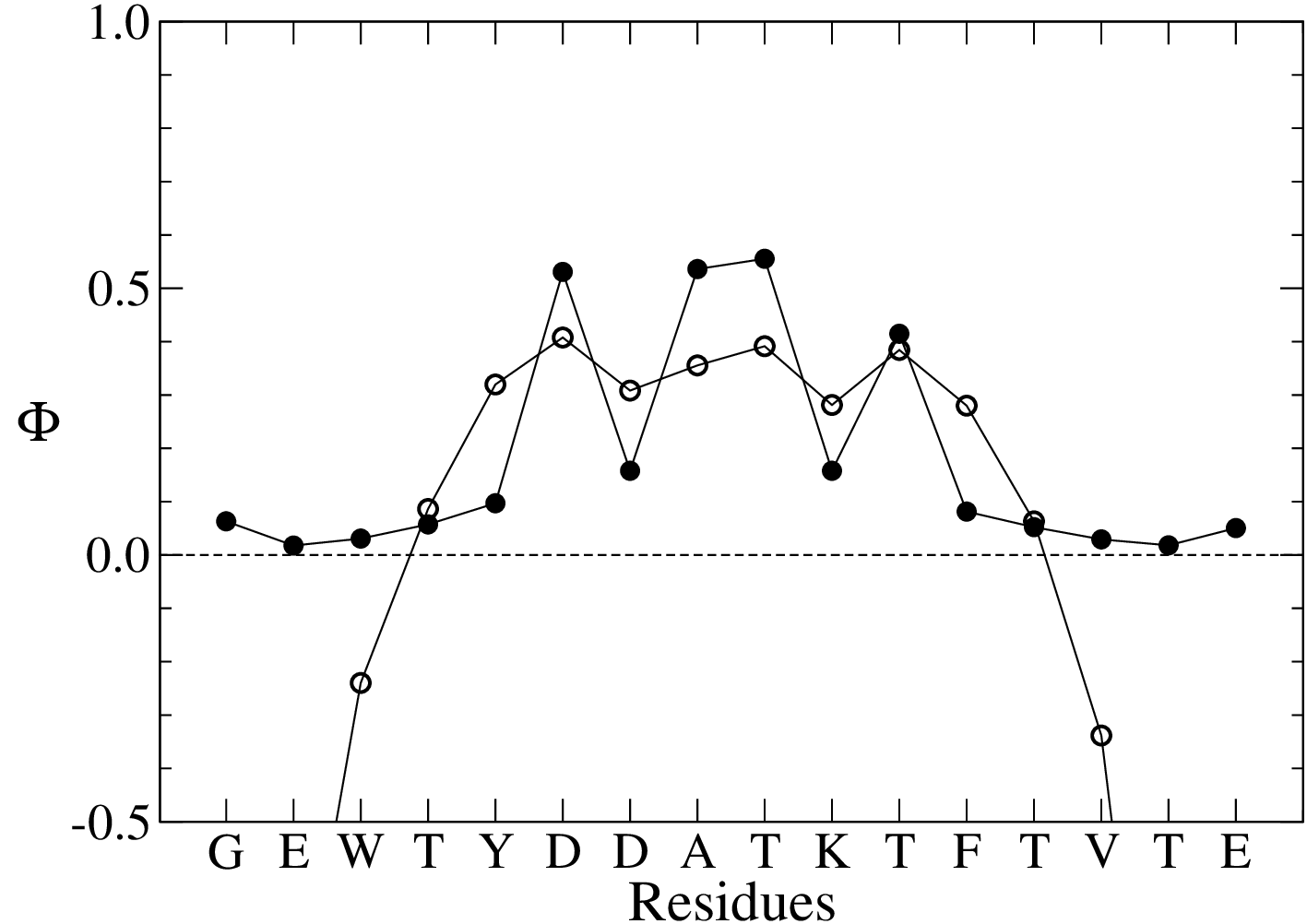}
\caption{
\label{fig:phival} 
Effects of ``mutations'' as measured by 
$\Phi-$values on each residue. Full circles: $\Phi-$values  
from the exact solution; open circles: $\Phi-$values
within MFA3 approach. The temperatures are in both cases those whereby
$F_U = F_N$. Results depend only slightly on temperature, anyway. 
}
\end{figure}

To evaluate the  $\Phi$-values,
we compute the variations in free energy profiles induced by each mutation,
for the exact solution and MFA3. 
$F_U$ and $F_N$ are evaluated as $F_{U,N}
=-RT\ln Z_{U,N}$, where $ Z_U$ $, Z_N$ are, respectively, the
partition functions 
restricted to unfolded and native basins in the free-energy profile,
i.e. the regions to the left and right of the top of the barrier in
Fig.~\ref{fig:profiles}. $F_\ddag$ is the free-energy of the top of the 
barrier.
Through expression Eq.~(\ref{eq:phi}) we obtain the $\Phi-$values for each
residue. In Fig.~\ref{fig:phival} we report the $\Phi-$value distributions.
There is a good overall correlation between the profiles, that
increase and decrease together. 
This is a further confirmation that the relevant features
of the model are conserved when applying the MF approach.
Mean-field results yield smoother profiles, as it could be expected. 
The ends of the hairpin are
characterized by low $\Phi$-values, that become negative for MFA3:
this would correspond to mutations that increase the stability gap but
decrease the barrier, or vice-versa. 
According to these results, the folding nucleus would be made up by
residues 6, 8, 9, 11, which is in contrast with the already mentioned
simulations.

\section{Conclusions} 
In this work we developed and discussed three different
mean-field schemes for the Galzitskaya-Finkelstein model, that represent
valid ways to deal with the model for characterizing the thermodynamical 
properties of a protein and its folding pathway as well.
These approaches offer viable alternatives both to the procedure proposed
by Galzitskaya and Finkelstein,\cite{Finkel} and to MC
simulations, that become computationally demanding for long polymers and
usually affected from the sampling problems.
We applied the model to the $\beta$-hairpin fragment 41-56 of the
GB 1 protein,  since, for this simple system, mean field results can be
compared with a brute force solution of the model, and both can be 
checked against experimental data and simulation published by other 
groups.

Our results suggest that, as far as specific heat and simple
thermodynamic quantities are concerned, the standard mean-field MFA1
is enough to yield correct results, provided that one uses the recipe
Eq.~(\ref{eq:matching}) to connect the two branches of the
solution. For more sophisticated quantities like free-energy profiles,
correlations and $\Phi$-values, MFA3 is to be preferred, since it
correctly recovers the main features of the exact solution. 
The hope is that mean-field results are still representative of the
exact ones in the case of longer and more complex proteins, where the
latter cannot be evaluated.

GF model itself yields results that are  somewhat in 
contrast with the MC and MD simulations on more detailed models 
for the hairpin. This discrepancy is probably due to the extreme 
simplicity of the hamiltonian Eq.~(\ref{eq:finkel}), where 
no distinction is made among the different kinds of interactions, 
such as hydrogen-bonds, side chain hydrophobicity, and so on. 
Indeed, we expected that a model accounting just for the
topology of the native state will not score very well when applied to
the $\beta$-hairpin, where detailed sequence information is
relevant.\cite{Koba2000} 
Predictions of the model could possibly improve if these elements were
taken into account.

\begin{acknowledgments}
We thank A.~Maritan, C.~Micheletti, A.~Flammini and A.~Pelizzola
for their suggestions and useful discussions about the model. 
F.C.~ thanks A.~Vulpiani and U.M.B.~Marconi and acknowledges the 
financial support of Cofin Murst 2001 on 
"Fisica Statistica di Sistemi Classici e Quantistici". P.B.~
acknowledges the financial support of Cofin Murst 2001 ``Approccio
Meccanico-Statistico ai Biopolimeri''.
\end{acknowledgments}



\begin{thebibliography}{99}
\bibitem{karpnote} Indeed, this is what Karplus and coworkers suggest in
                   Ref.~{\protect{\onlinecite{Karplus}}}; yet other results
                  \protect{\cite{Pande, Klimov}} do not support this view.
\end{thebibliography}
\end{document}